\documentclass[aps,nofootinbib,twocolumn,notitlepage,floatfix,superscriptaddress,secnumarabic]{revtex4-1}

\usepackage{indentfirst}
\usepackage{graphicx,color}
\usepackage{multirow}
\usepackage{times}
\usepackage{bm}
\usepackage{tabularx}
\usepackage{epstopdf}
\usepackage{enumerate}
\usepackage[normalem]{ulem}
\usepackage{color}
\usepackage{float}
\usepackage[colorlinks, linkcolor=red,
            anchorcolor=blue, citecolor=green]{hyperref}
\usepackage[T1]{fontenc}

\begin{document}

\title{
Determination of impact parameter in high-energy heavy-ion collisions via deep learning}
\author{Pei Xiang}
\affiliation{Department of Physics and Center for Field Theory and Particle Physics, Fudan University, Shanghai, 200433, China}

\author{Yuan-Sheng Zhao}
\affiliation{Department of Physics and Center for Field Theory and Particle Physics, Fudan University, Shanghai, 200433, China}

\author{Xu-Guang Huang}
\email{huangxuguang@fudan.edu.cn}
\affiliation{Department of Physics and Center for Field Theory and Particle Physics, Fudan University, Shanghai, 200433, China}

\affiliation{Key Laboratory of Nuclear Physics and Ion-beam Application (MOE), Fudan University, Shanghai 200433, China}

\date{\today}

\begin{abstract}
In this study, Au+Au collisions with the impact parameter of $0 \leq b \leq 12.5$ fm at $\sqrt{s_{NN}} = 200$ GeV are simulated by the AMPT model to provide the preliminary final-state information. After transforming these information into appropriate input data (the energy spectra of final-state charged hadrons), we construct a deep neural network (DNN) and a convolutional neural network (CNN) to connect final-state observables with impact parameters. The results show that both the DNN and CNN can reconstruct the impact parameters with a mean absolute error about $0.4$ fm with CNN behaving slightly better. 
Then, we test the neural networks for different beam energies and pseudorapidity ranges in this task. It turns out that these two models work well for both low and high energies. But when making test for a larger pseudorapidity window, we observe that the CNN shows higher prediction accuracy than the DNN. With the method of Grad-CAM, we shed light on the `attention' mechanism of the CNN model.
\end{abstract}
\maketitle

\section{Introduction}\label{intro}
\setlength{\parindent}{2em} As the unique means to generate quark-gluon plasma (QGP) on earth, high-energy heavy-ion collision experiments provide us opportunities to study this kind of extremely hot and dense matter. With more research on the collective behaviour of quarks and gluons, both the deep structure of a nucleus and the state of the universe at a few microsecond after the Big Bang have been brought to light. For heavy-ion collisions, besides the collision energy, the impact parameter (denoted by $b$) is another crucial quantity which determines the initial geometry of a collision. Numerous quantities have essential correlations with the impact parameter. For example, the elliptic flow of hadrons is very sensitive to the impact parameter \cite{Heinz:2013th}. The electromagnetic (EM) fields in heavy-ion collisions roughly satisfy $e\vert {\rm Field} \vert \propto \sqrt{s} f(b/R_A)$ where $f(b/R_A)$ is a function of $b/R_A$ with $R_A$ the nucleus radius \cite{huang2016electromagnetic,Hattori:2016emy}. When studying the EM properties of QGP, dilepton production is a significant probe. For lepton pair production, not only the cross section but the azimuthal asymmetry has a strong dependence on impact parameter \cite{li2020impact}. The recently observed hyperon spin polarization increases with increasing impact parameter of the collisions \cite{Liu:2020ymh,Gao:2020vbh}. However, the impact parameter of a single collision cannot be measured directly in heavy-ion experiments. Usually, it is estimated by particular final-state observables sensitive to it, such as the charged-particle multiplicity. By introducing the concept of centrality, which is defined as classes classified by $b$, and comparing experimental data with simulation results by, e.g., the Glauber model, one can determine the rough interval of the impact parameter of an event \cite{abelev2013centrality}.

Due to the powerful ability to establish a reliable map between the input data and the target value without that much prior knowledge, deep learning (DL) methods are widely used in not only science research, but also our daily life. When applying DL algorithms to face recognition tasks, a machine can identify one's ID with his/her facial information \cite{balaban2015deep}. For heavy ion collisions, the impact parameter can be viewed as one of the IDs of an event. Several works have proved the effectiveness of DL methods on impact parameter `recognition' \cite{david1995impact, bass1996neural, haddad1997impact, li2020application, kuttan2020fast,Li:2021plq,Tsang:2021rku,Zhang:2021zxd,Mallick:2021wop}. From a simple neural network \cite{david1995impact} to a PointNet model \cite{kuttan2020fast} and to boosted decision trees \cite{Mallick:2021wop}, with the development of machine learning algorithms, more and more appropriate learning methods have been proposed to improve the performance of `recognition' and to satisfy experimental requests. However, most of these researches only involve collisions at low or intermediate energies \cite{david1995impact, bass1996neural, haddad1997impact, li2020application, kuttan2020fast,Li:2021plq,Tsang:2021rku,Zhang:2021zxd}. Though a recent work \cite{Mallick:2021wop} considered LHC energies, the adopted machine learning model is not a deep neural network. 

Here, we investigate RHIC Au+Au collisions at $\sqrt{s_{NN}} = 200$ GeV and choose final-state charged hadrons as probes. After transforming these particles' momentum information into energy spectra as input data of learning models, we use a deep neural network (DNN) model and a convolutional neural network (CNN) model, respectively, to find a map between the energy spectrum and the impact parameter. Then, we analyze the influences of beam energy and the range of pseudorapidity on the accuracy of predictions in this task. Furthermore, we examine the interpretability of the CNN serving as a regression machine. An `attention' map of the CNN model is obtained by the Grad-Cam algorithm. All of the collisions are simulated by A Multi-Phase Transport (AMPT) model \cite{lin2005multiphase}.

\section{Deep Learning Algorithms}\label{dl}
The discovery of Higgs Boson in 2012 completes the jigsaw of elementary particles according to the current Standard Model. Not long from that, CERN announced the Higgs boson machine learning challenge \cite{adam2015higgs} to the public. The goal of this challenge is to help experimentalists to distinguish the signal of Higgs boson decay from background noise better by machine learning (ML) methods. It turned out that ML was extremely effective in this task.

As a branch of the field of artificial intelligence, ML technology has promoted intelligentization in many areas of industry, such as autonomous driving, smart mobile electronic devices, internet industry, and so on. Due to the same request of dealing with a large amount of data or obtaining information which is hard to fetch by traditional methods, scientists have applied ML techniques to fundamental science research, and physics is no exception. On the whole, the applications of ML in physics can be divided into two categories. One is the replacement of physical models with ML models if the latter ones are more effective for specific problems. By training ML models with a big amount of data, one can construct a mapping between two or more physical quantities. For instance, by training neural networks to imitate wave functions approximately, we can construct a map between the potential and one particle's energy without solving the Schr\"{o}dinger equation \cite{mills2017deep}, or solve the quantum many-body problem \cite{carleo2017}. The other one is to recognize the target signal (such as a physical phenomenon) from the background with noise. For example, deep neural networks can help distinguish Higgs boson or other exotic particles of interest (signal) from other particles (background) \cite{baldi2014searching}. In short, ML algorithms can be used to deal with regression and classification problems in physics. As a branch of ML, DL  becomes the most popular AI method in physics research. In the field of relativistic heavy ion collision, DL has been applied to the problems of QCD phase transition \cite{Du:2019civ,pang2018equation,Wang:2020tgb}, relativistic hydrodynamics \cite{huang2019applications}, study of jet structure \cite{komiske2017deep,Du:2021pqa}, the search of chiral magnetic effect \cite{Zhao:2021yjo}, recognition of initial clustering structure in nuclei \cite{He:2021uko}, etc. Among various DL algorithms, the deep neural network and the convolutional neural network are two common models.

\begin{figure}[h]
	\centering
	\includegraphics[height=4cm,width=8cm]{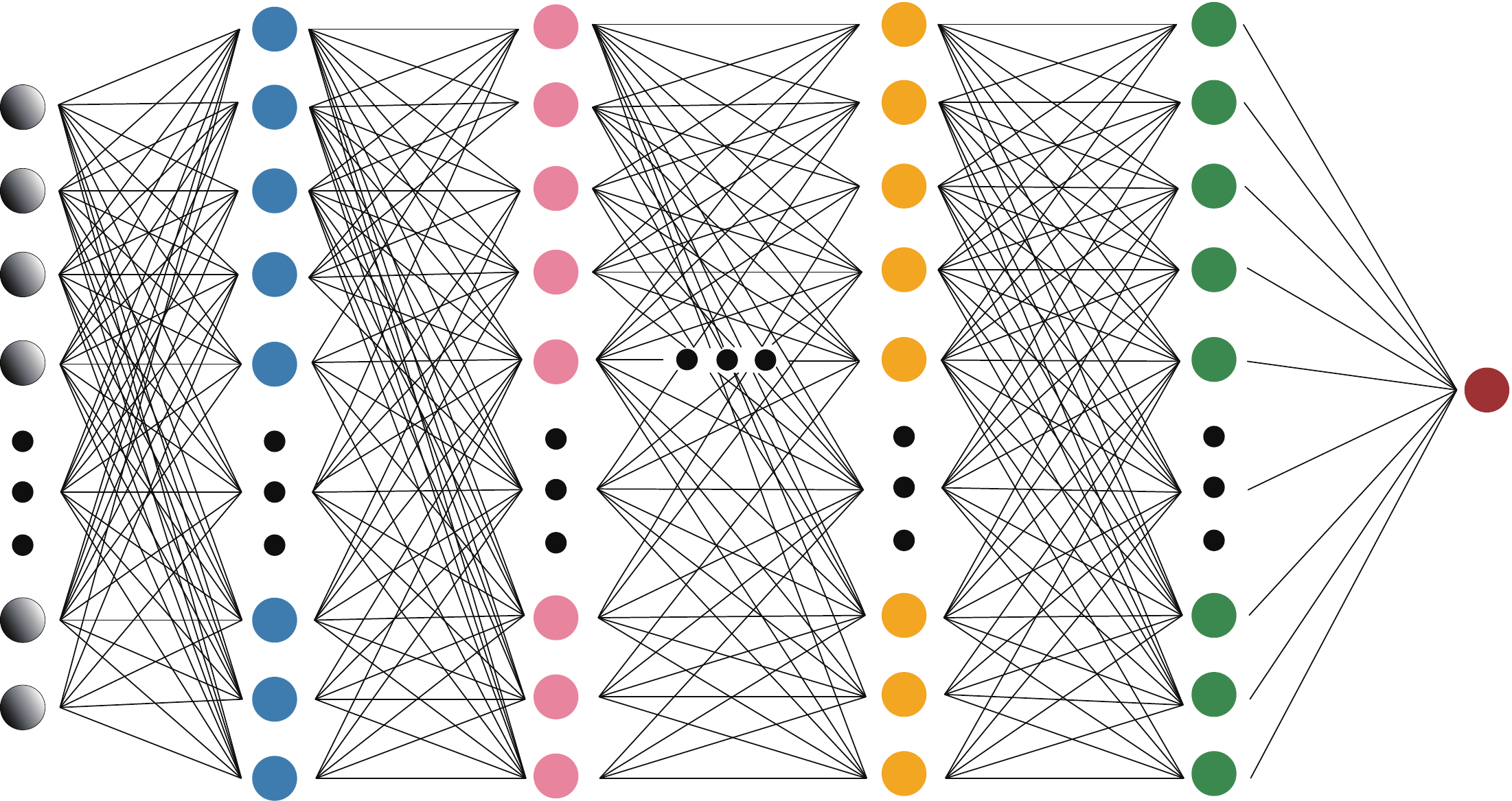}
	\caption{The architecture of the deep neural network (DNN) we use in this work. It contains 4 hidden layers. The 1st layer is composed of 512 neurons and there are 256 neurons in the 2nd layer, 128 neurons in the 3rd one, 64 neurons in the last hidden layer.}
	\label{p1}
\end{figure}

The deep neural network (DNN) is one of the early DL models. Due to its remarkable ability to realize nonlinear mapping with a comparatively simple structure, the DNN is still a preferred tentative model for most regression tasks. A DNN (Fig.\ref{p1}), also known as multilayer perceptron, is composed of an input layer, enough hidden layers to be `deep', and an output layer.

\begin{figure}[h]
	\centering
	\includegraphics[width=0.5\textwidth]{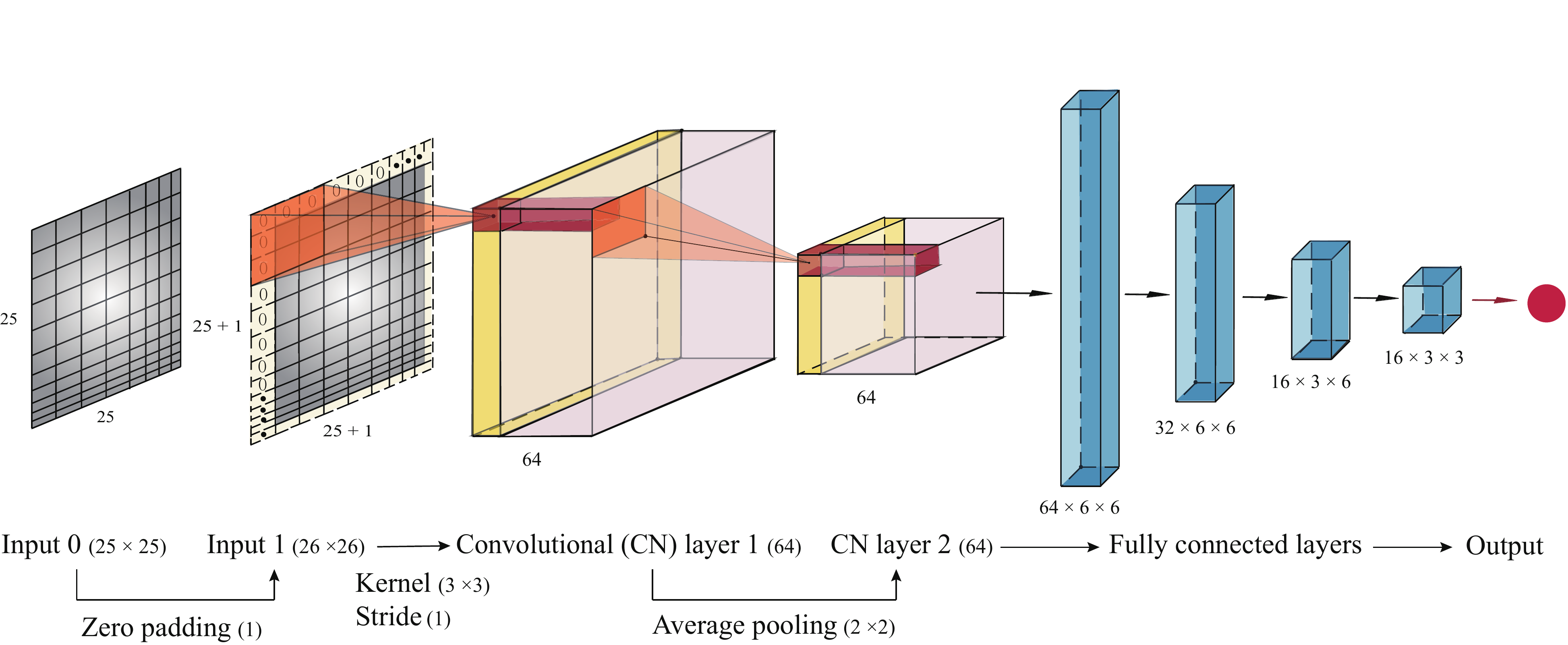}
	\caption{The structure of the convolutional neural network (CNN). It contains 2 convolutional layers and 4 fully-connected layers. }
	\label{p2}
\end{figure}

The convolutional neural network (CNN) commonly appears in 2D image-related tasks. A typical CNN consists of the input layer, convolutional layers, pooling layers, fully connected layers, and the output layer. Our CNN architecture is shown in Fig.\ref{p2}.

As a supervised learning regression task, impact parameter determination aims at constructing a mapping between the input observables and a single value, i.e. the impact parameter of an event. Thus, it is appropriate to choose the mean squared error (MSE) as the loss function serving as examining the performance of the learning models. It is defined as
\begin{equation}
{\rm Loss} = \frac{1}{N_{batch}} \sum_{i=1}^{N_{batch}}(y_i^{pred} - \hat{y}_i^{true})^2 ,
\end{equation}
where $\hat{y}_i^{true}$ is the true value of impact parameter of an event among a batch of events whose size is $N_{batch}$, and $y_i^{pred}$ is the output of the DNN/CNN model as corresponding prediction value.

\section{Simulation of events and Selection of Observables}\label{ampt}
We consider Au+Au collisions at $\sqrt{s_{NN}}=200$ GeV and use A Multi-Phase Transport (AMPT) model to perform the simulation. The AMPT model \cite{lin2005multiphase} is a hybrid transport model which contains four basic stages:
the initial condition, partonic scattering, hadronization, and hadronic interaction. The initial condition is generated by HIJING model \cite{wang1991hijing}. Scatterings among partons are modeled by Zhang's Parton Cascade (ZPC) model \cite{Zhang:1997ej}. Differing from each other in the process of hadronization, two versions of AMPT model are available. One is the default model in which partons are recombined with their parent strings and the Lund string fragmentation model is used to turn partons into hadrons. The other one contains a string melting model. It combines partons into hadrons via a quark coalescence model. Finally, the rescatterings of the hadronic matter are performed by A Relativistic Transport (ART) \cite{Li:1995pra}. Here, we choose the one with string melting as the simulator.

The selection of features serving as input information is the first step of a deep learning process. Considering the experimental observability, we choose momenta of the freezed-out charged hadrons with pseudorapidity $\eta$ satisfying $-1 \leq \eta \leq 1$ as observables.

\begin{figure}[h]
	\centering
	\includegraphics[width=0.5\textwidth]{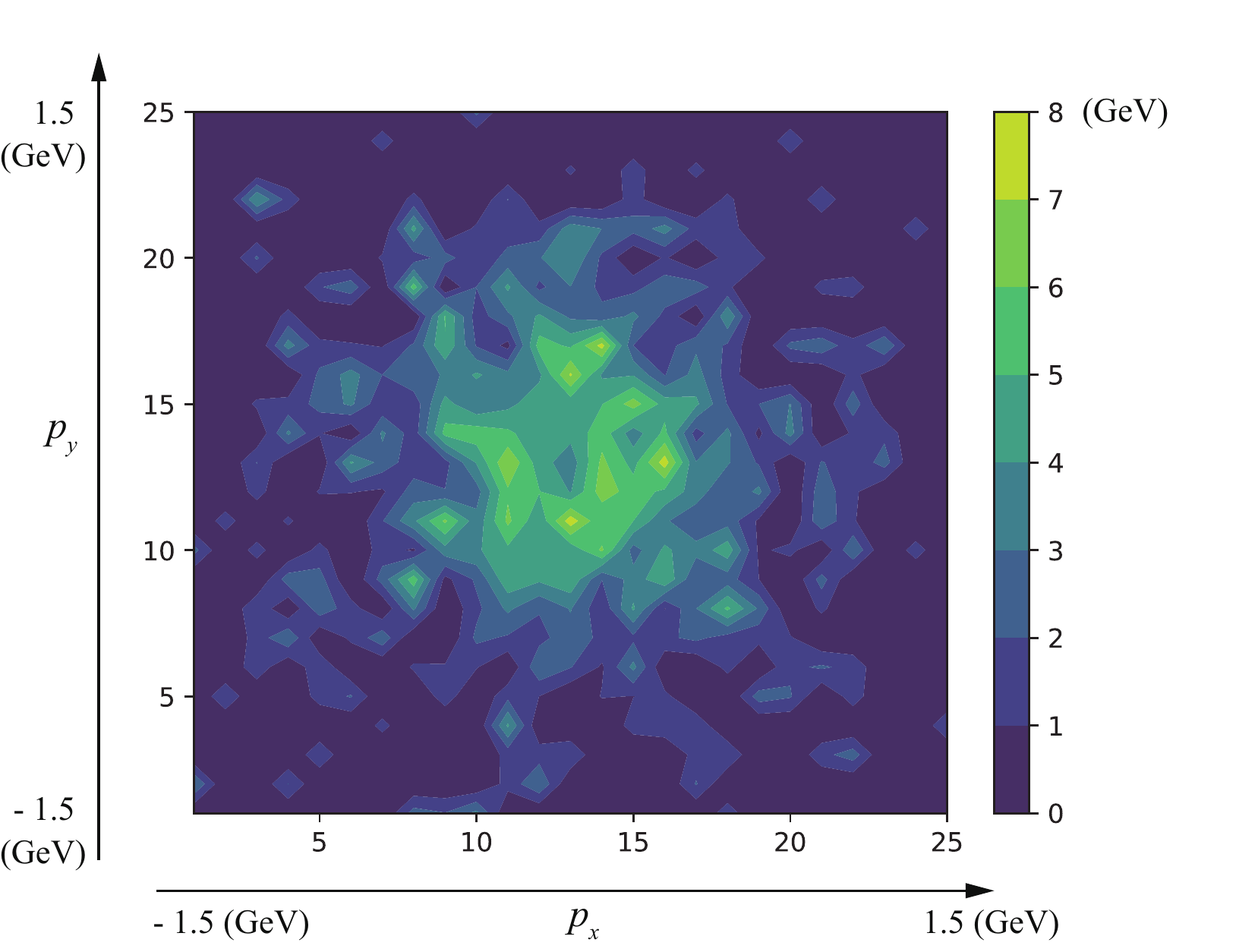}
	\caption{The energy spectrum in ($p_x$, $p_y$) space of final-state charged particles of an event with $b=1.65$ fm. It is cut by a $25 \times 25$ grid. Every unit in the grid represents the sum of energies of the enclosed particles. This kind of energy spectrum is used as the input data fed to DL models.}
	\label{p3}
\end{figure}

As shown in Sec. \ref{dl}, the input data of a DNN model and that of a CNN model have different forms. Here, for the CNN model, the observables above are transformed into 2-dimensional energy spectra in ($p_x$, $p_y$) space. Fig.\ref{p3} illustrates the case of an event with $b=1.65$ fm. The $x, y$ components of momenta are cut by the interval $[-1.5, 1.5]$ and the ($p_x$, $p_y$) space are cut by a $25 \times 25$ grid. Every cell in this grid contains the sum of energies of charged particles whose $p_x$ and $p_y$ are within corresponding intervals. Note that the energy contains not only the momentum but also the mass of the particle. Thus, an energy spectrum carries more physical information than a multiplicity spectrum. Arrange the pixels in a 2D energy spectrum into an 1D chain, the data is generated as an input for the DNN model.

\section{Results and Discussion}
We generate 28,000 events per centrality and split them into two parts: 20,000 events for training and 8,000 events for validation. The interval of impact parameter $ b \in [0, 12.5]$ fm is divided to 9 centrality classes \cite{qiu2011event}. Thus, the total training dataset is composed of 180,000 events and the total validation dataset contains 72,000 events. After training the model into an appropriate one, 20,000 events (satisfying a differential distribution $\propto b \textit{$d$}b$ in impact parameter) are fed to the model to test its prediction accuracy.

\begin{figure}[h]
	\centering
	\includegraphics[width=0.4\textwidth]{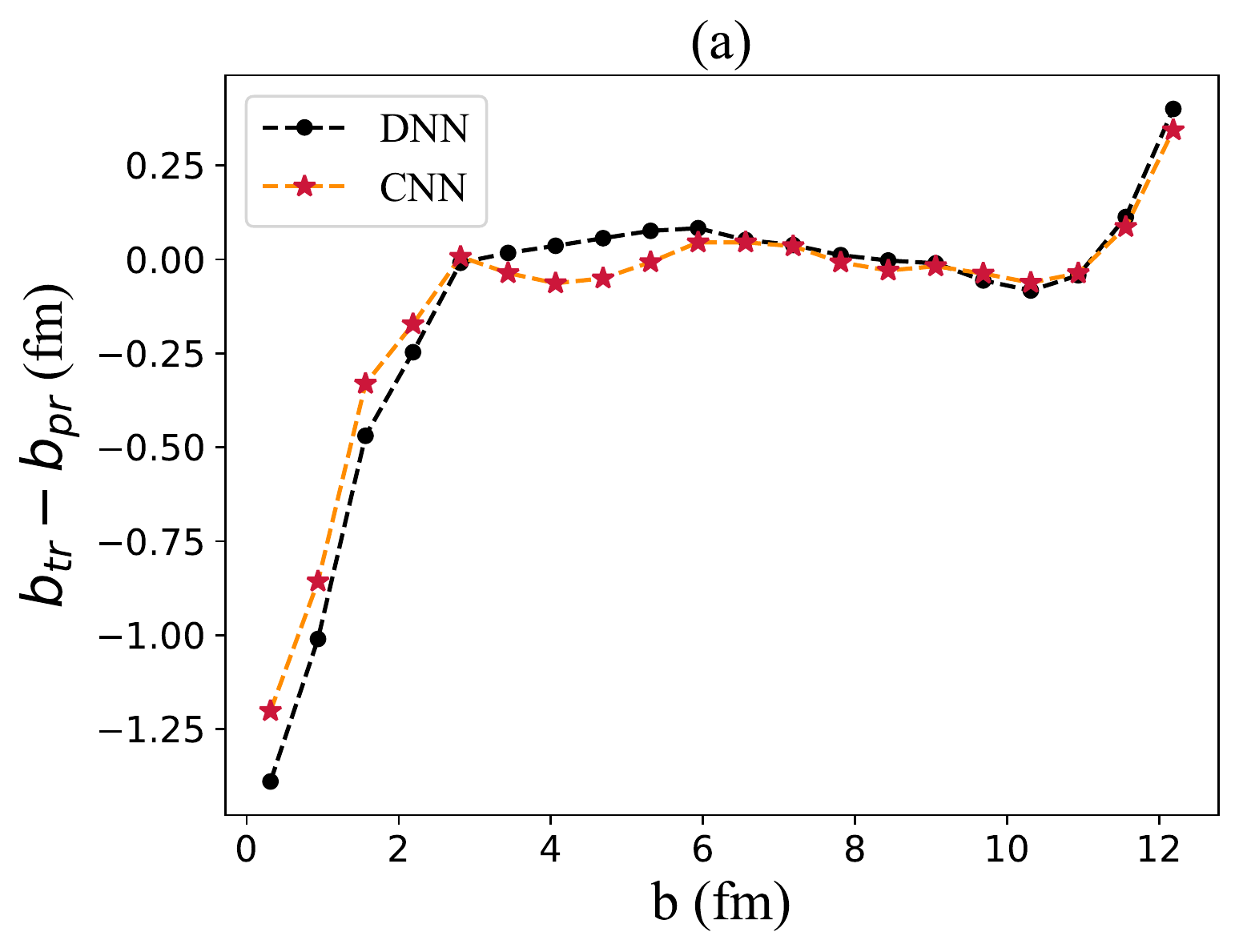}
	\includegraphics[width=0.4\textwidth]{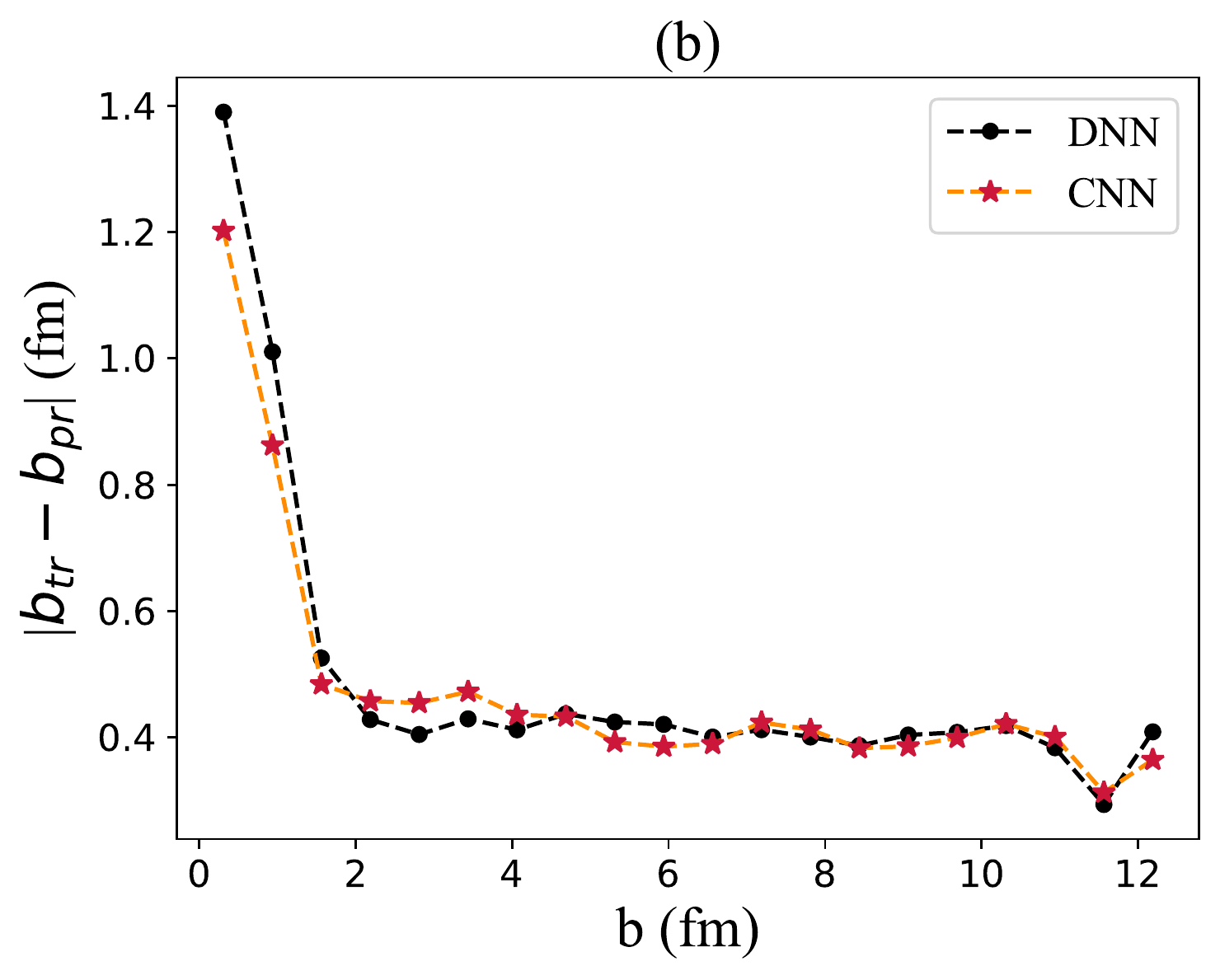}
	\caption{Errors between true values of impact parameter and those predicted by the DNN / CNN model for Au+Au collisions at $\sqrt{S_{NN}} = 200$ GeV: (a) The mean errors between true values (denoted by $b_{tr}$) and predicted values (denoted by $b_{pr}$); (b) Corresponding mean absolute errors.}
	\label{p4}
\end{figure}

In Fig.\ref{p4}, we show the errors of the predicted impact parameters comparing to the true values. Both the DNN and CNN models show high prediction accuracy for semi-central and semi-peripheral events. The mean error of the CNN model for events with impact parameter satisfying $2$ fm $\leq b \leq 11$ fm ranges from $-0.06$ fm to $0.05$ fm and that of the DNN model ranges from $-0.08$ fm to $0.08$ fm. 
The mean absolute prediction error is $0.40$ fm for the CNN and $0.41$ fm for the DNN models, showing that the CNN model performs slightly better than DNN model. But the accuracy decreases in central regions Prediction values are greater than true values for central collisions and the case of peripheral ones is opposite.

\subsection*{A. Robustness Over Collision Energies}
\begin{figure}[h]
	\centering
	\includegraphics[width=0.4\textwidth]{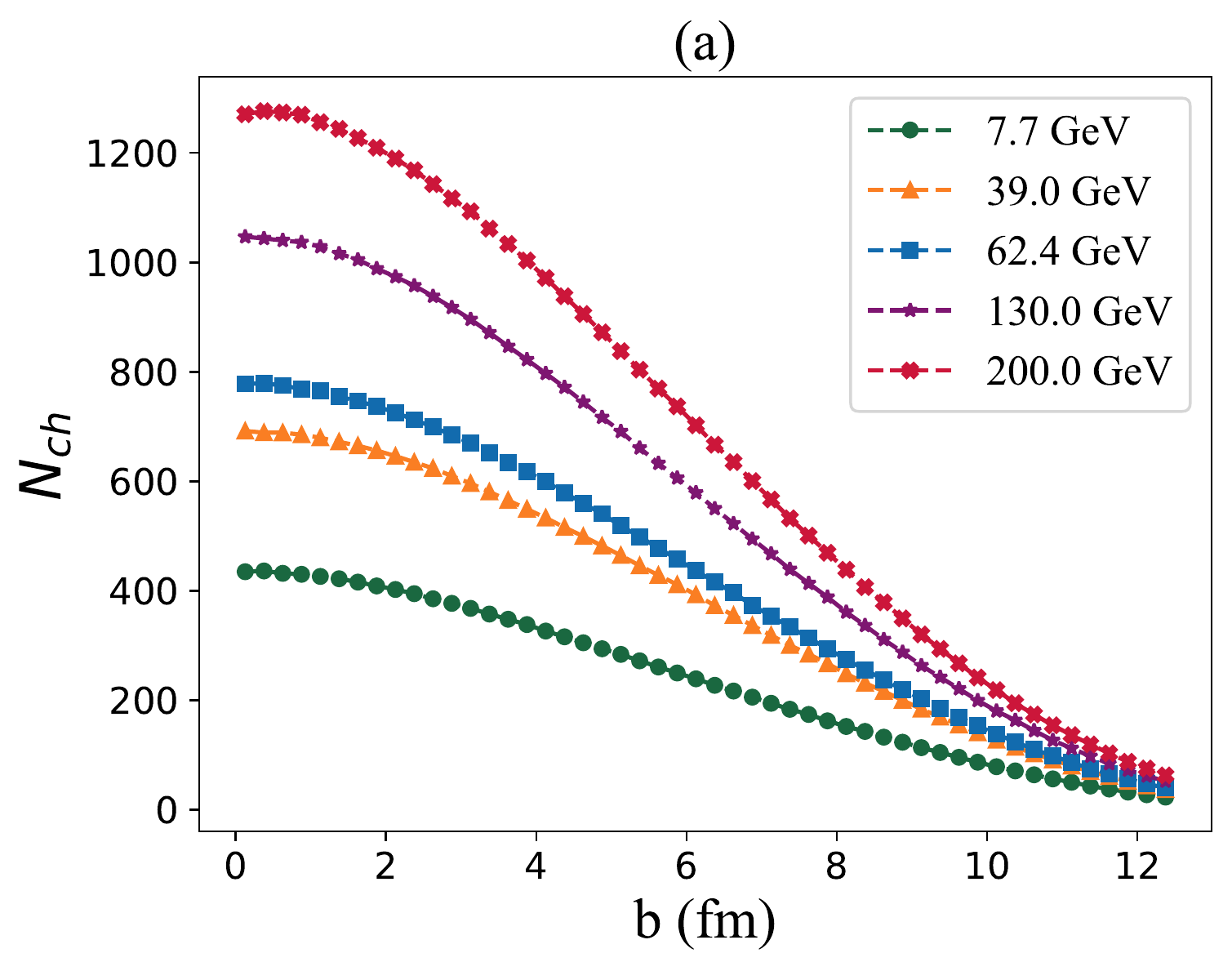}
	\includegraphics[width=0.4\textwidth]{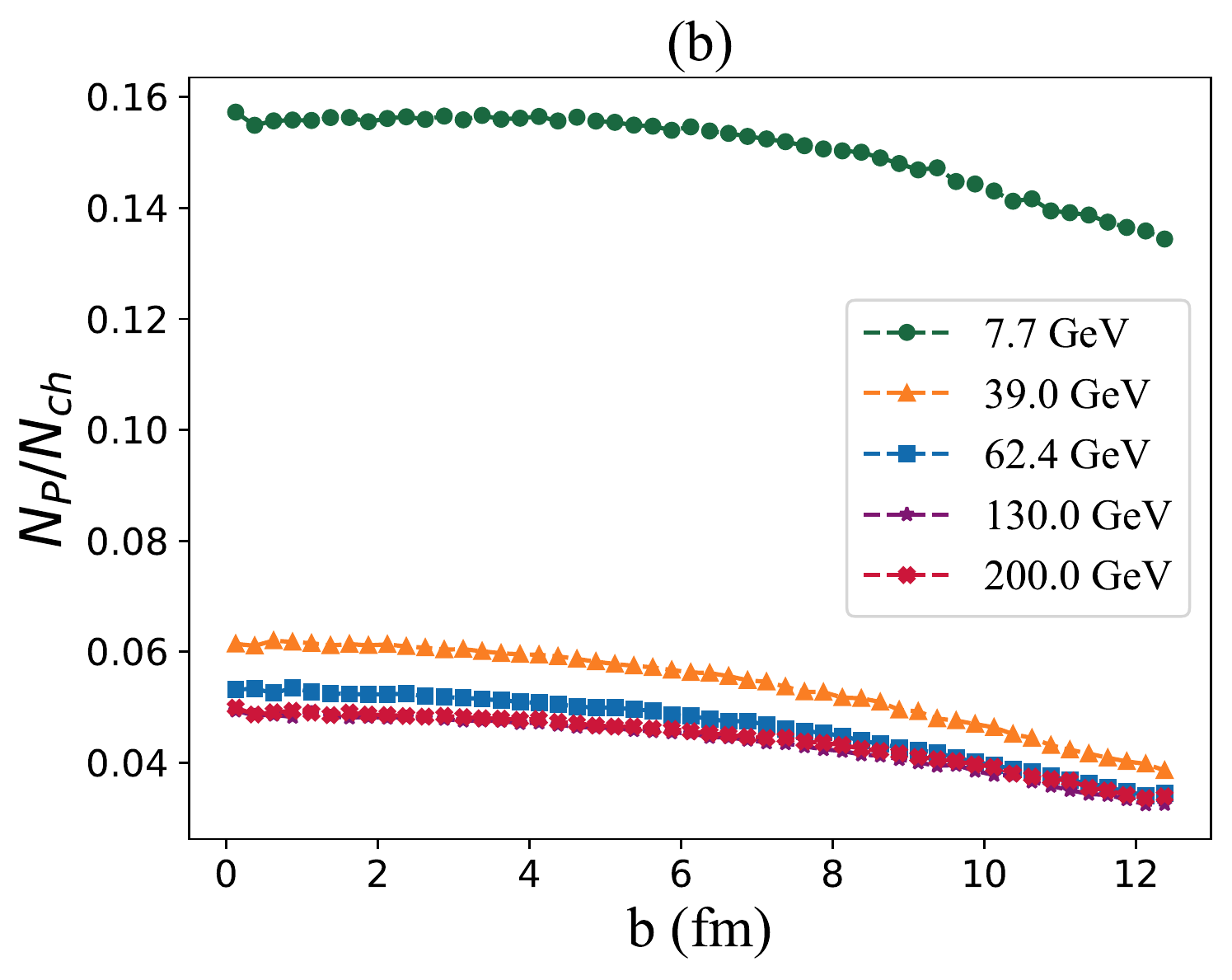}
	\includegraphics[width=0.4\textwidth]{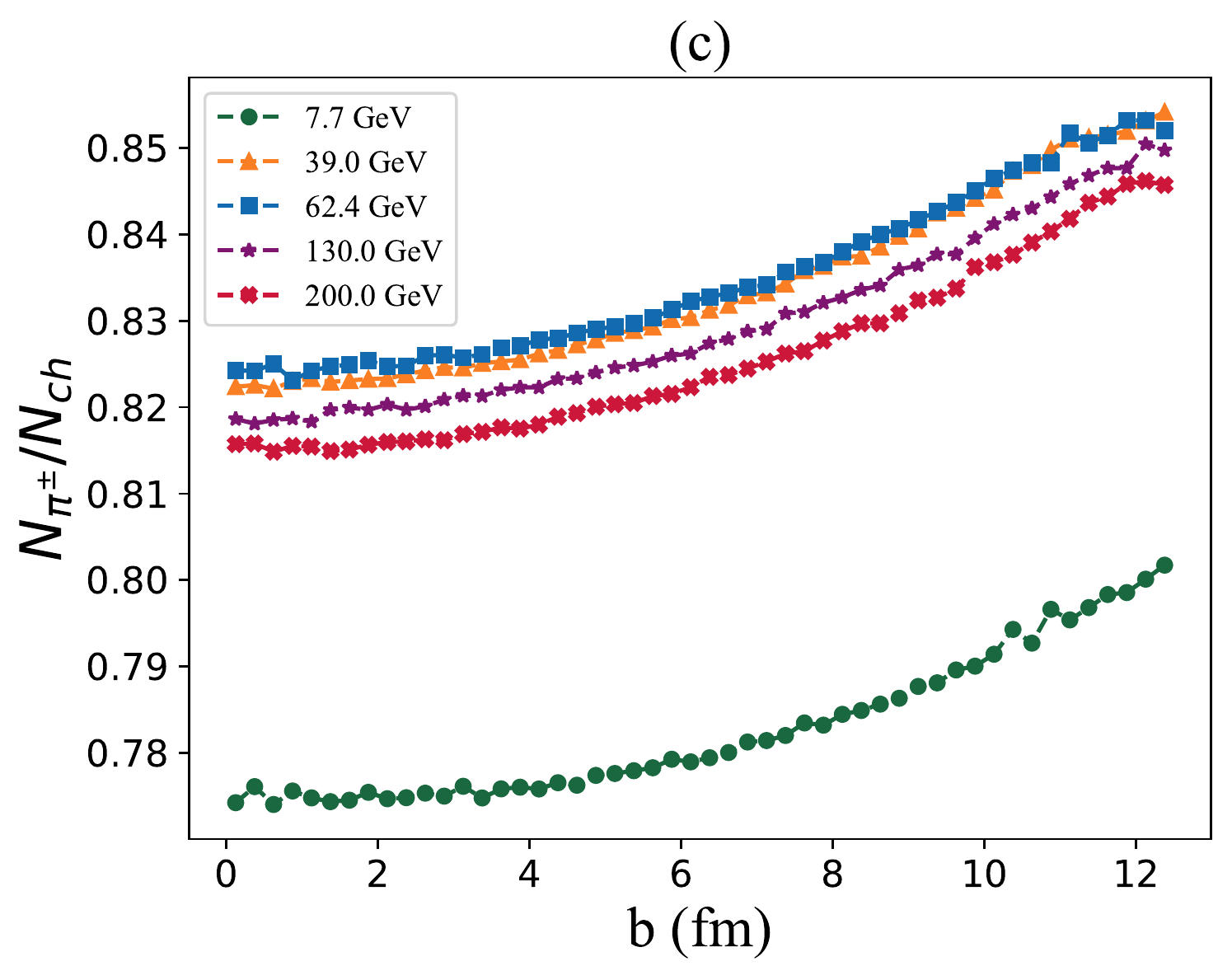}
	\caption{The multiplicity and composition of charged particles produced in collisions with different beam energies, i.e. $\sqrt{s_{NN}}$ = 7.7, 39.0, 62.4, 130.0, 200.0 GeV: (a) The multiplicity of charged particles; (b) The fraction of protons in charged particles; (c) The fraction of charged pions.}
	\label{p5}
\end{figure}

In heavy ion collisions, the beam energy is another crucial quantity which largely determines the bulk properties of the matter created in an event. Low energy collisions are different from high energy ones in many aspects. 
With higher beam energy, colliding nuclei generate more partons and finally more varieties of hadrons freeze out. To illustrate this, we generate dataset for events of collision energies $\sqrt{s_{NN}} = 7.7,\ 39.0,\ 62.4,\ 130.0,\ 200.0$ GeV corresponding to the beam energy scan program performed at RHIC. Then we analyze their differences in multiplicity and composition of final state charged particles. As shown in Fig.\ref{p5}, the multiplicity of charged particles increases with the beam energy. In addition, the fraction of protons in charged hadrons is higher at lower collision energy but the fraction of charged pions is not monotonic in collision energy. We thus test whether the differences in multiplicity and composition of produced hadrons at lower collision energies affect the ability of the DL models.

\begin{figure}[h]
	\centering
	\includegraphics[width=0.4\textwidth]{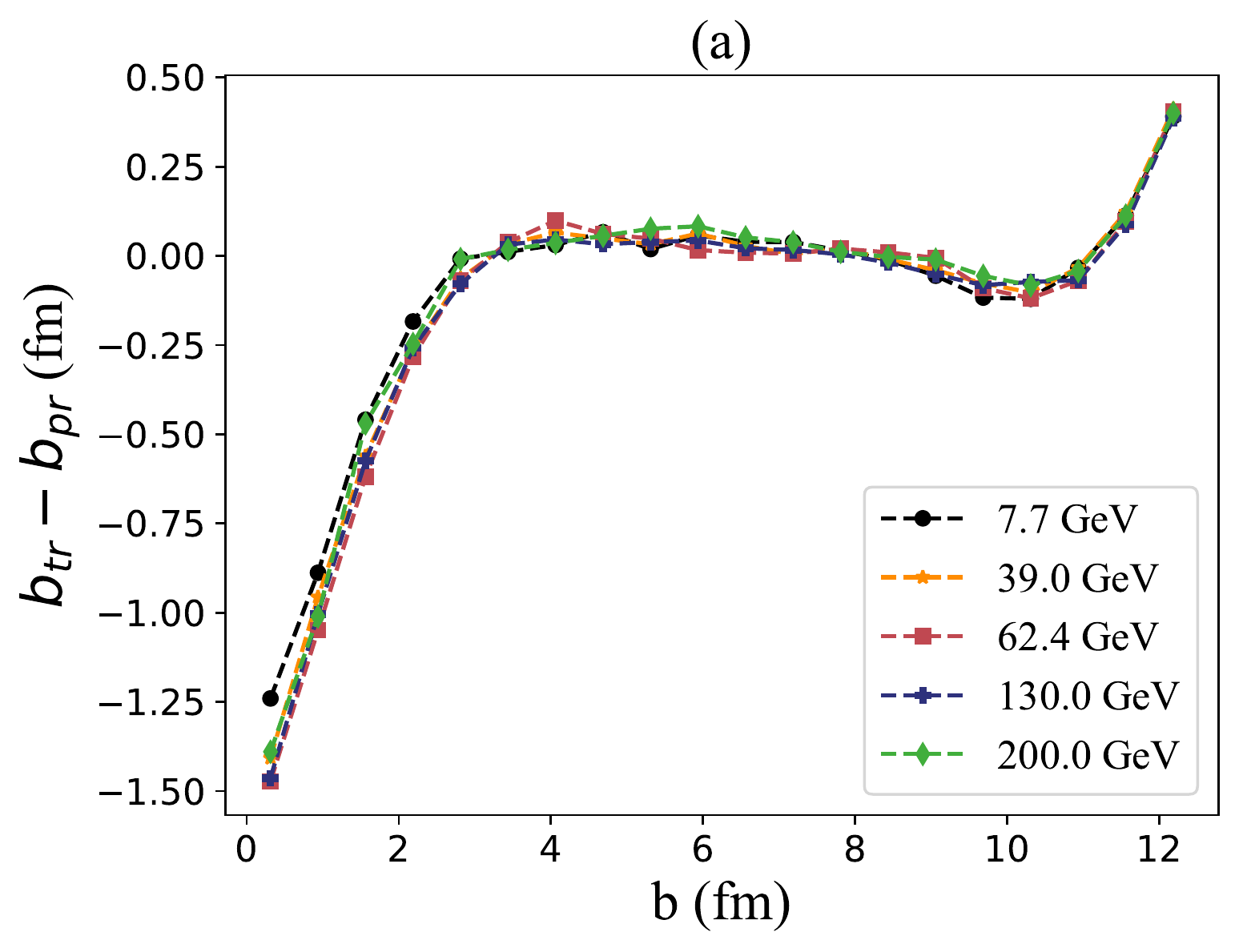}
	\includegraphics[width=0.4\textwidth]{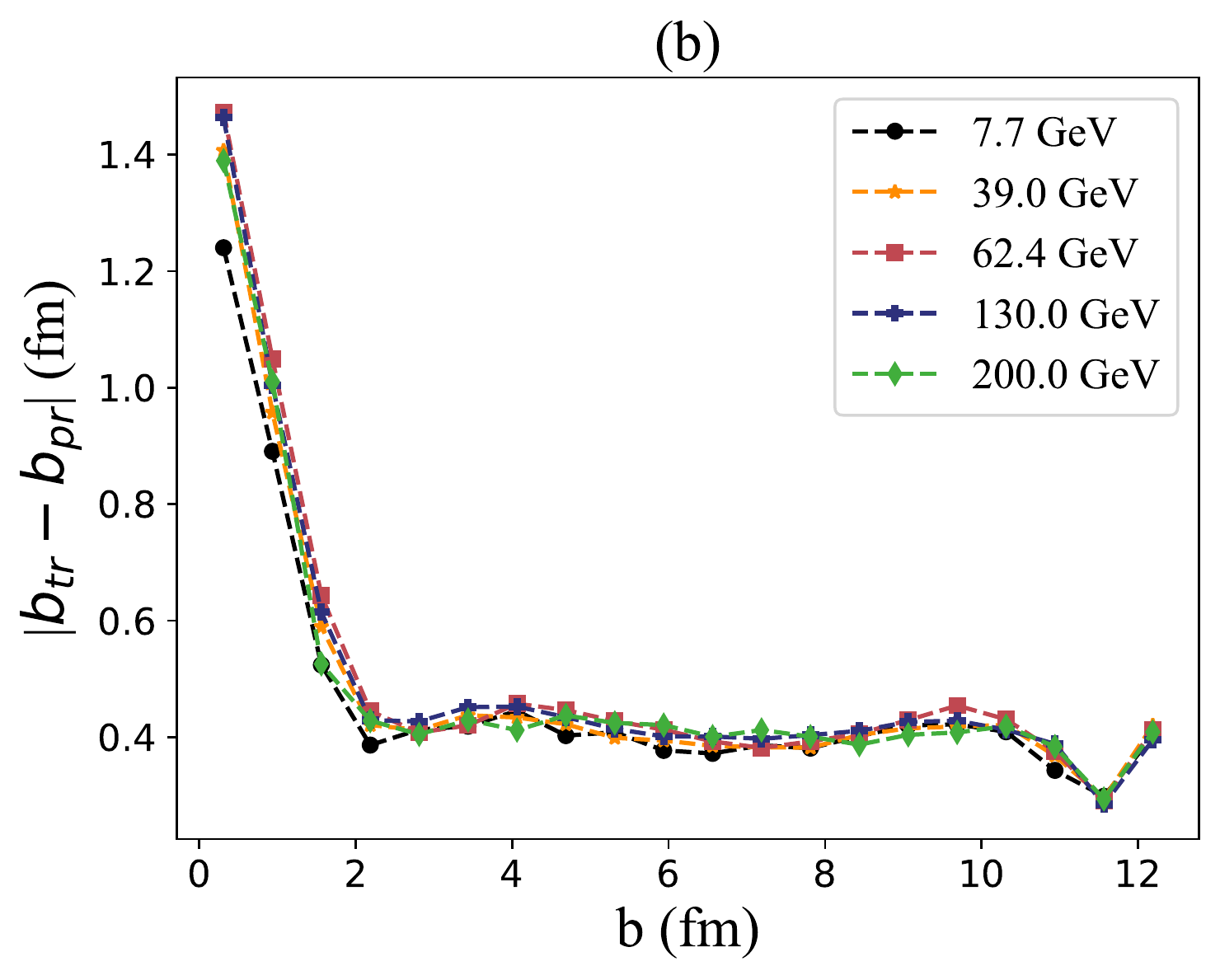}
	\caption{Performance of the DNN in dealing with collisions with different beam energies, i.e. $\sqrt{s_{NN}}$ = 7.7, 39.0, 62.4, 130.0, 200.0 GeV: (a) The mean errors between true values (denoted by $b_{tr}$) and predicted values (denoted by $b_{pr}$); (b) Corresponding mean absolute errors.}
	\label{p6}
\end{figure}

\begin{figure}[h]
	\centering
	\includegraphics[width=0.4\textwidth]{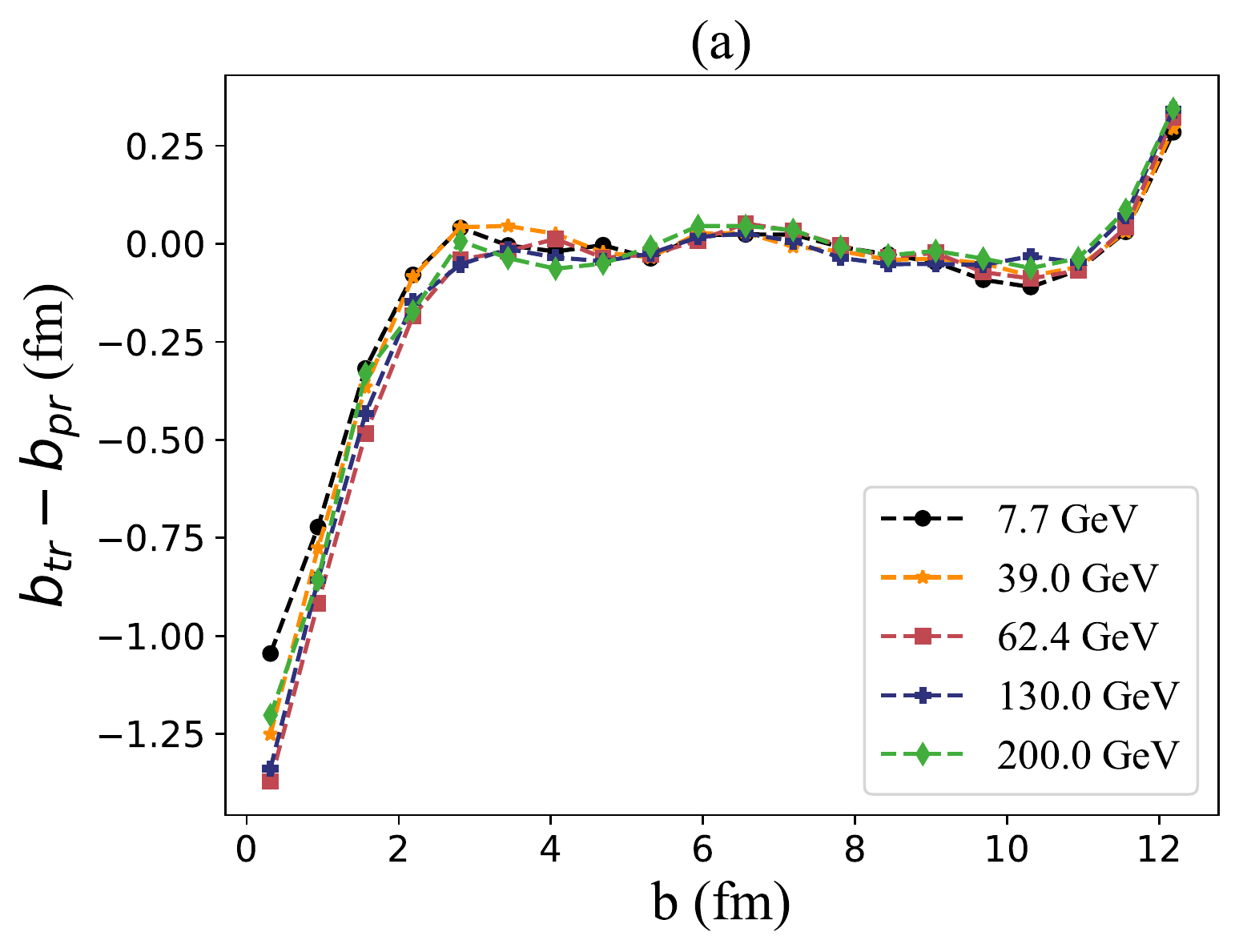}
	\includegraphics[width=0.4\textwidth]{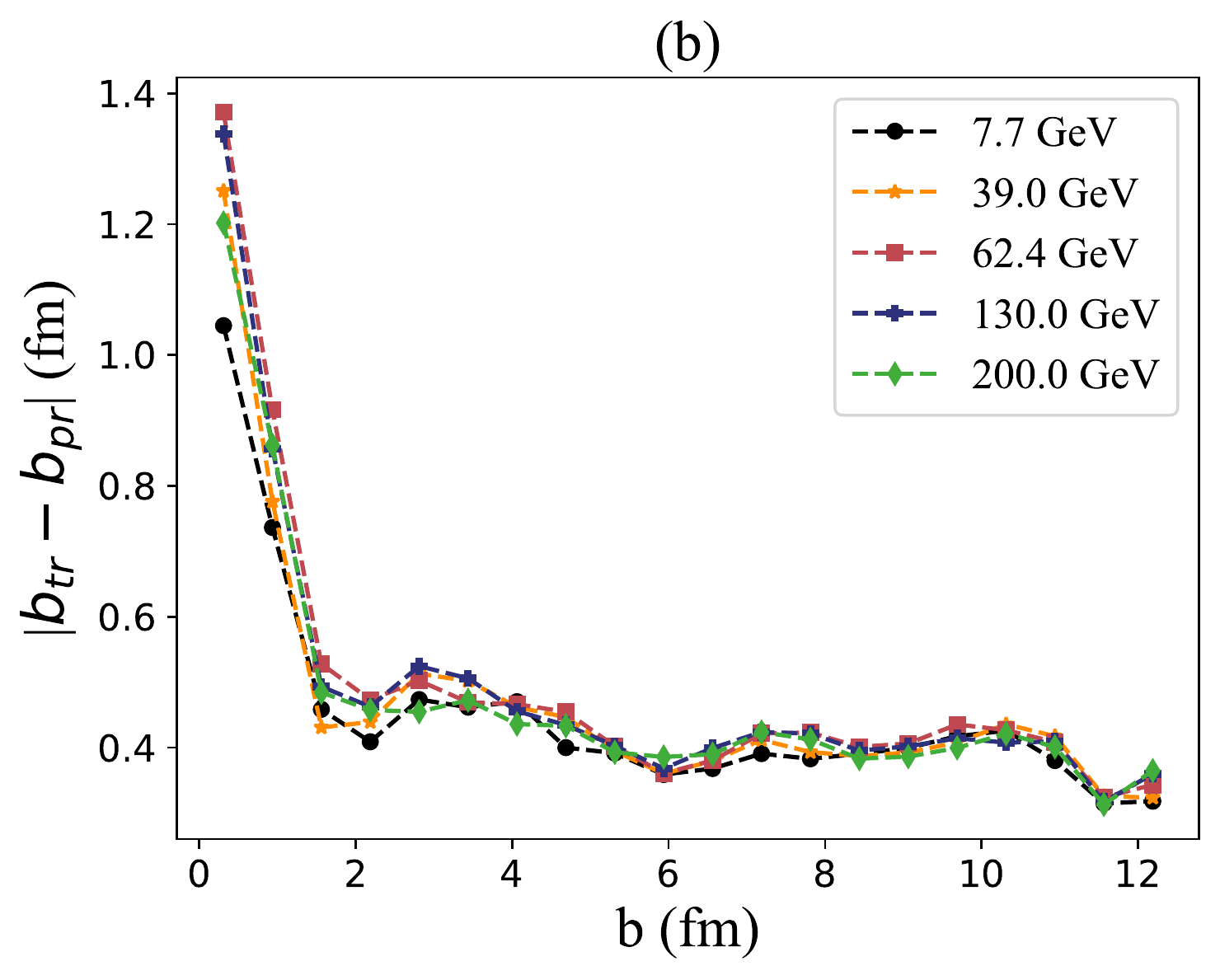}
	\caption{Performance of the CNN in dealing with collisions with different beam energies, i.e. $\sqrt{s_{NN}}$ = 7.7, 39.0, 62.4, 130.0, 200.0 GeV: (a) The mean errors between true values (denoted by $b_{tr}$) and predicted values (denoted by $b_{pr}$); (b) Corresponding mean absolute errors.}
	\label{p7}
\end{figure}

Therefore, we train and test the DL models to the cases of $\sqrt{s_{NN}} = 7.7,\ 39.0,\ 62.4,\ 130.0, \ 200.0$ GeV. The results are shown in Fig.\ref{p6} and Fig.\ref{p7} for DNN model and CNN model, respectively. The DL models perform well for low and intermediate energy cases as well. Thus we find that the DL models are very robust against different collisions energies.

\subsection*{B. Influence of Pseudorapidity Cut}
\begin{figure}[h]
	\centering
	\includegraphics[height=7cm,width=9cm]{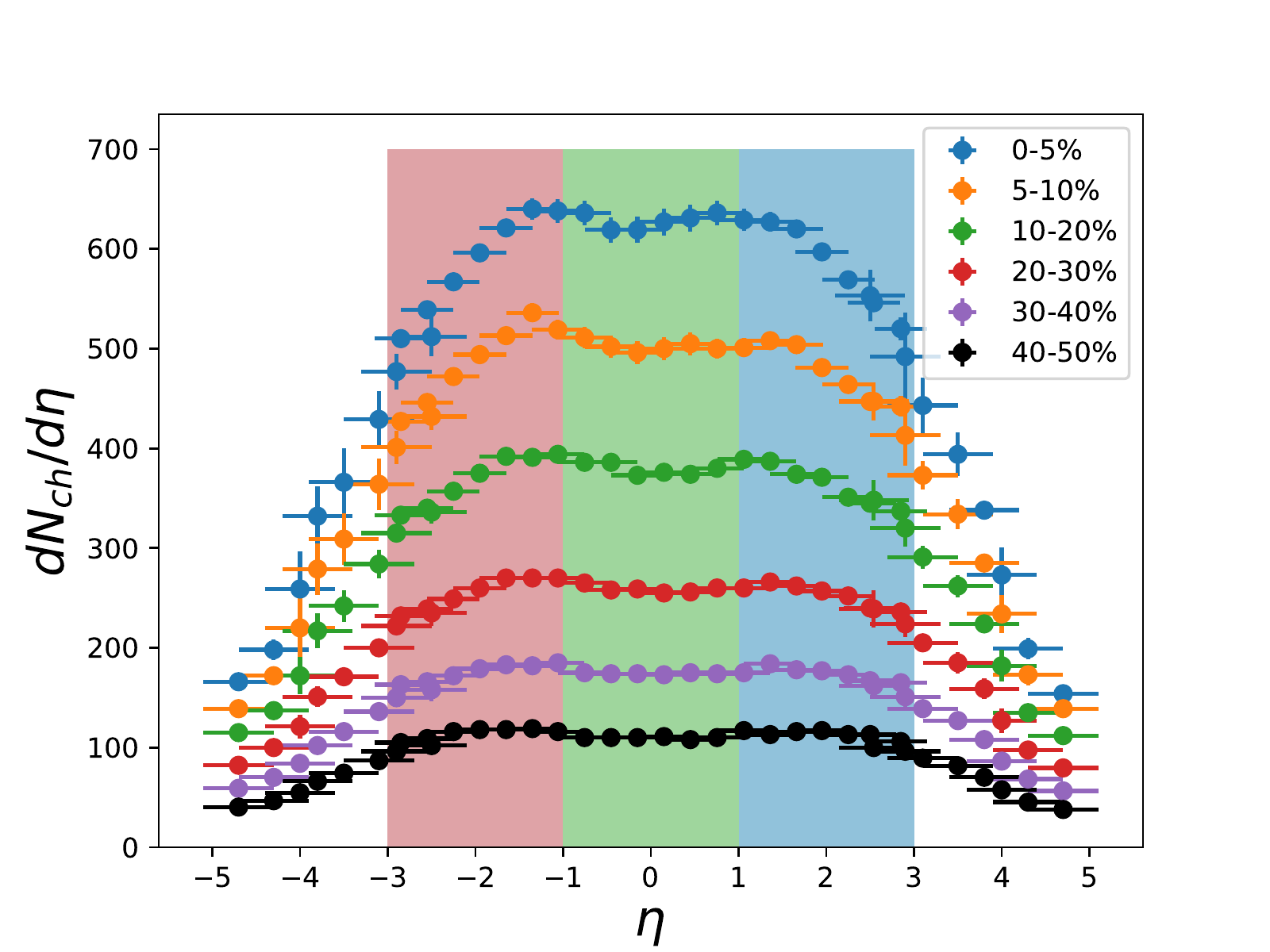}
	\caption{Distributions of $dN_{ch} / d\eta$ for centrality ranges of, top to bottom, (0-5) \%, (5-10) \%, (10-20) \%, (20-30) \%, (30-40) \%, and (40-50) \% \cite{bearden2002pseudorapidity}. 
The particles in the pseudorapidity window of $-3 \leq \eta \leq 3$ are split into 3 parts, i.e. [-3, -1], [-1, 1], and [1, 3], corresponding to 3 channels, `R', `G', and `B' of the input layer.}
	\label{p8}
\end{figure}

In most of relativistic heavy ion collision experiments, the greatest attention is focused on the mid-rapidity region, i.e. $-1 \leq \eta \leq 1 $, due to coverage limit of most of the detectors. Actually, the mid-rapidity region just covers a small part of final-state charged particles \cite{bearden2002pseudorapidity}, see Fig.\ref{p8}. Therefore, it is certain that the energy spectrum contains more information if we can observe a larger region in pseudorapidity. In fact, the octagon detector in the PHOBOS experiment can accept charged particles with $|\eta| < 3.2$ \cite{sarin2003measurement}, the recent upgrade of the inner Time Projection Chamber (iTPC) detector at RHIC can extend the rapidity acceptance to $-1.5 \leq \eta \leq 1.5 $ \cite{Luo:2017faz}.  Thus, we take the regions of $-3 \leq \eta \leq -1 $ and $1 \leq \eta \leq 3 $ into consideration.

\begin{figure}[h]
	\centering
	\includegraphics[width=0.4\textwidth]{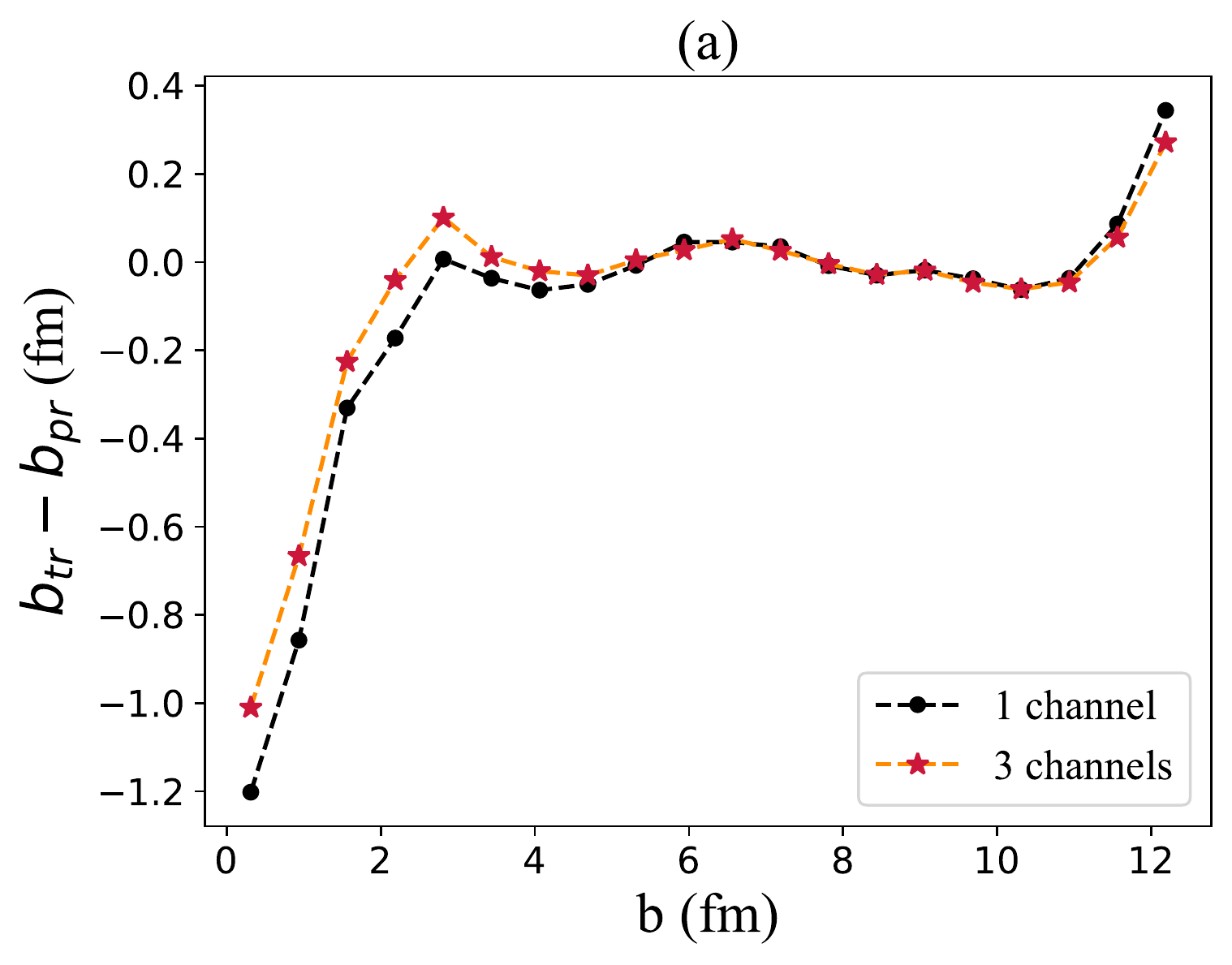}
	\includegraphics[width=0.4\textwidth]{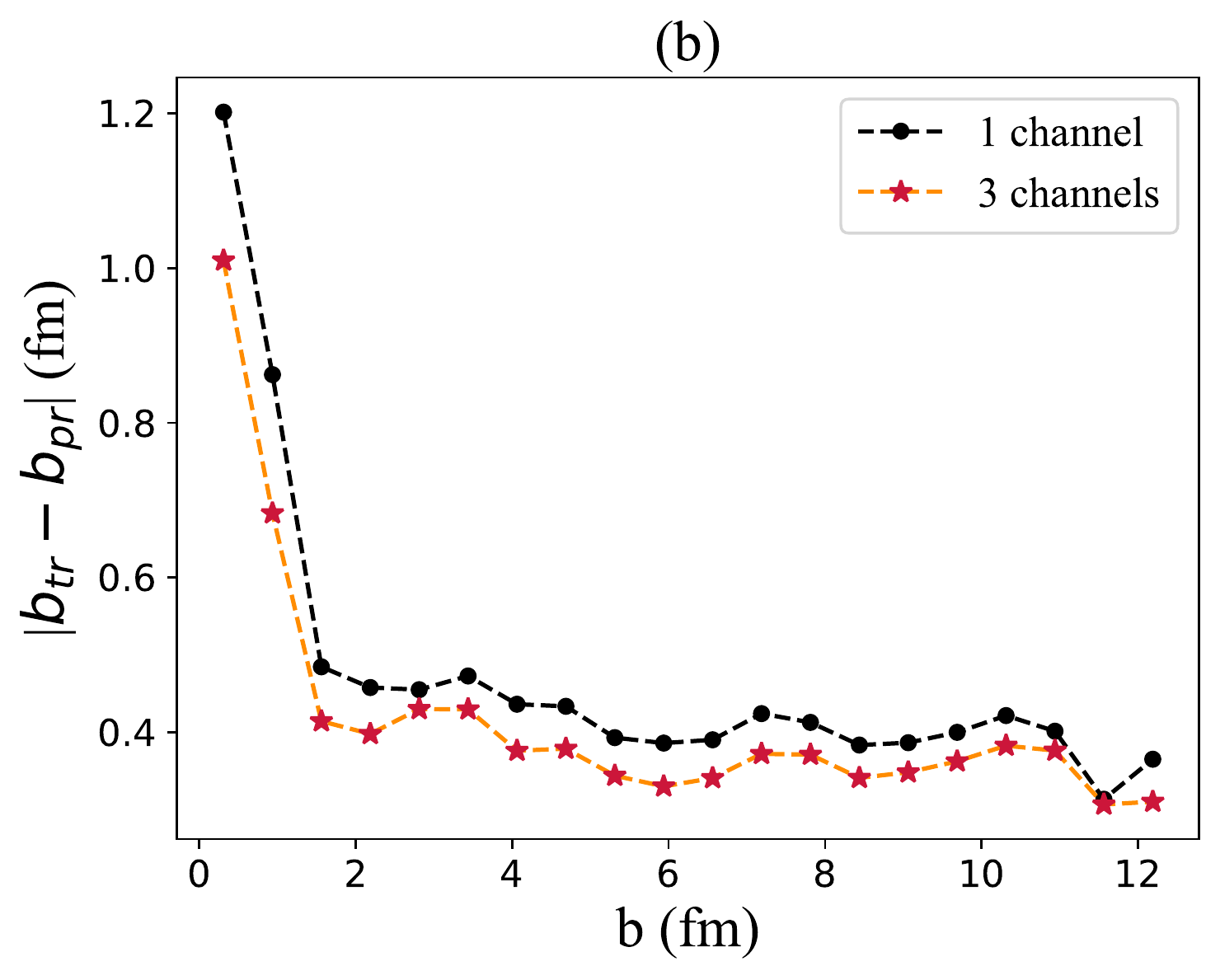}
	\caption{Errors between true values of impact parameter and those predicted by the CNN model with $\sqrt{S_{NN}} = 200$ GeV. The number of input channels is 1 and 3 respectively: (a) The mean errors between true values (denoted by $b_{tr}$) and predicted values (denoted by $b_{pr}$); (b) Corresponding mean absolute errors.}
	\label{p9}
\end{figure}

Now, we expand the training and detection region to $-3 \leq \eta \leq 3 $. Because the particles with positive pseudorapidity move in the opposite direction relative to those with negative pseudorapidity, we add two extra channels into the input layer of the CNN. Just like inputting a colored picture with 3 channels `RGB' into the CNN, we choose the region of $-3 \leq \eta \leq -1 $ as the channel `R', $-1 \leq \eta \leq 1 $ as the channel `G', and $1 \leq \eta \leq 3 $ as the channel `B' (Fig.\ref{p8}). Then, we feed the energy spectra with 3 channels above to the CNN we trained with only data in `G' region. The CNN shows higher prediction accuracy than the case of 1 channel (Fig.\ref{p9}). Now, the mean absolute prediction error of the CNN for $2 \leq b \leq 12.5$ fm is $0.35$ fm. This value is 0.40 $fm$ in the case of CNN with only 1 input channel. Thus, extending the pseudorapidity window can truly improve the performance of regression. 

\subsection*{C. Comparison with Multiplicity Method}
As we mentioned before, the impact parameter of a single event cannot be measured directly experimentally. And the situation is the same for geometrical quantities, such as the participant number $N_{part}$ and binary collision number $N_{coll}$. Instead, one can introduce the quantity `centrality', which is usually expressed as a percentage of the total nuclear interaction cross section \cite{abelev2013centrality} and strongly correlated with the impact parameter $b$, to estimate an event's $b$ range. In heavy ion collision experiments, usually the multiplicity of final state charged particles ($N_{ch}$) is chosen to be the main observable to classify events' centrality. It is assumed that the average multiplicity of charged particles decreases monotonically when the impact parameter increases. With the Glauber model and Monte Carlo method, we can establish the centrality classes for heavy ion collisions \cite{miller2007glauber}. By comparing the charged-particle multiplicity of an event measured by experiments with the results given by the MC-Glauber model, one can determine this event's centrality and further estimate it's impact parameter.

Here, we propose a scheme to compare the uncertainty of $b$ determined by multiplicity method with the CNN method. Based on the above assumption that the impact parameter is monotonically related to charged-particle multiplicity, if we pick out a batch of events with the same multiplicity ($N_{ch}$), their impact parameters are supposed to satisfy a Gaussian-like distribution. We select the standard deviation ($\sigma_{b}^{mult}$) of these impact parameters as the quantity to characterize the uncertainty of mapping $N_{ch}$ to the corresponding impact parameter, which can be taken as the center of the Gaussian-like distribution, i.e. $\mu_{b}^{mult}$. Then, a batch of events with the same impact parameter $\mu_{b}^{mult}$ are generated by the AMPT model. With a well-trained CNN model, we will obtain corresponding predicted impact parameters and calculate their standard deviation, i.e. $\sigma_{b}^{CNN}$. By comparing $\sigma_{b}^{mult}$ with $\sigma_{b}^{CNN}$, we can evaluate the uncertainty of impact parameter determination by multiplicity method or by CNN method.

\begin{figure}[h]
	\centering
	\includegraphics[width=0.5\textwidth]{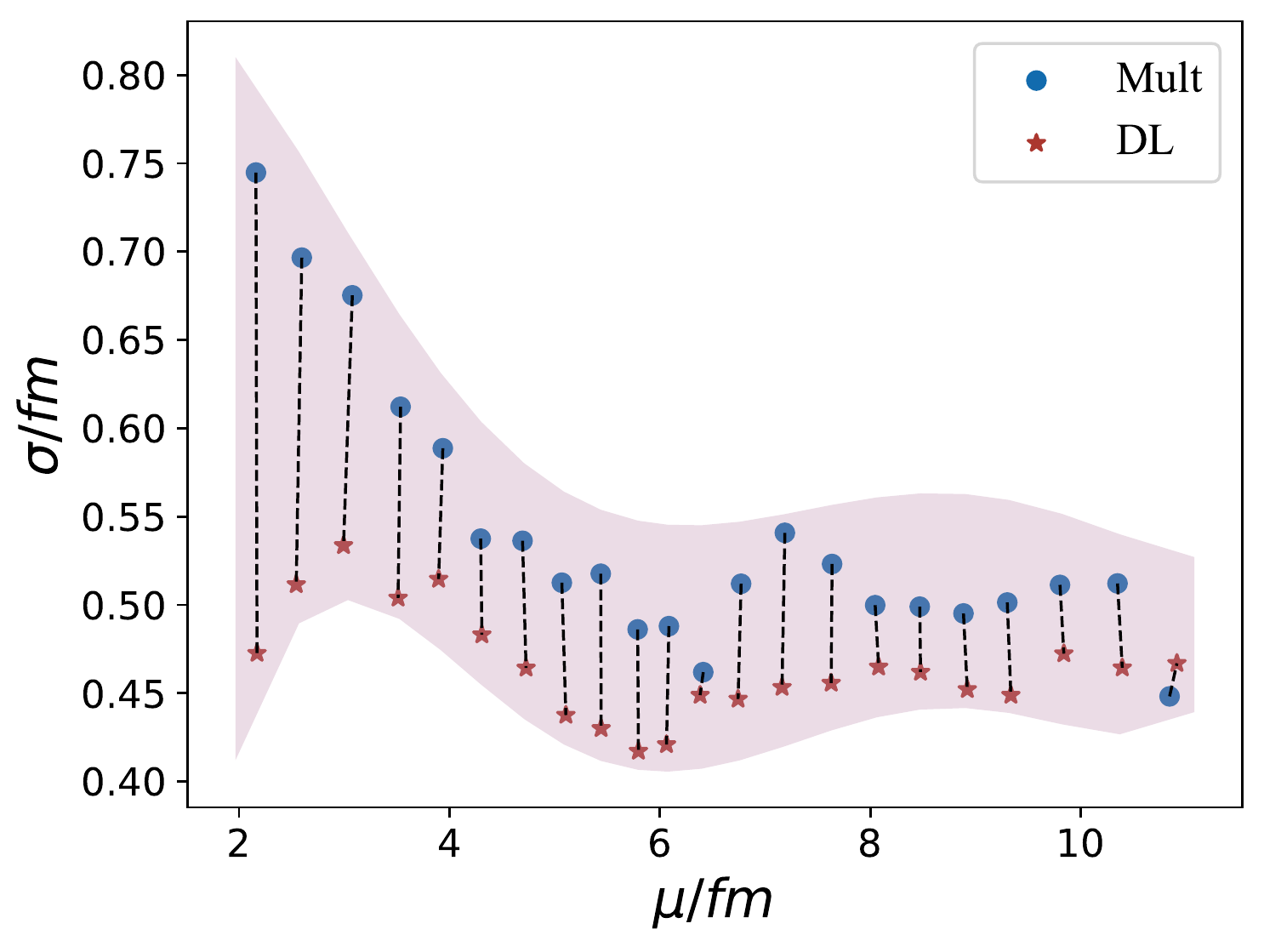}
	\caption{Comparing the standard deviation of $b$ distribution given by multiplicity method with that given by the CNN. $\mu$ is the mean value of the $b$ distribution. `Mult' means the multiplicity method and `DL' refers to the deep learning method, which is the CNN model here.}
	\label{p10}
\end{figure}

Due to the event-by-event fluctuation in nucleon distribution in the nuclei before the collision and in the complex evolution process after the collision, events with the same impact parameter ought to have different charged-particle multiplicities and different energy spectra. Thus, evaluating the uncertainty of these two methods above is equivalent to measuring the degree of single-valued correspondence between impact parameter and the multiplicity or the energy spectrum. From the results shown in Fig.\ref{p10}, it can be concluded that the charged-particle energy spectrum with CNN as identifier of impact parameter behaves better than the multiplicity method.

\section{Fetching Physical Information from the CNN}
Deep learning algorithms have shown their strong capability to construct a map between the input data and the target. As a result, we can succeed in various regression or classification tasks without prior knowledge. However, most of the widely used DL models are not interpretable. Due to their `complexity' and `dimensionality', it is hard to understand how these models work and fetch instructive information from them \cite{bathaee2017artificial}. So these DL models are viewed as `black boxes' in most cases. More and more effort has been made to open the `black boxes' of DL algorithms. By analyzing the DNNs in the Information Plane \cite{tishby2015deep}, one can have an understanding of the training and learning processes and furthermore the internal representations of the DNNs \cite{shwartz2017opening}. In addition, for CNNs which are usually used in visual recognition tasks, there has been a number of works aimed at visualizing them. By operating global average pooling and defining a quantity measuring the importance of neurons in a CNN, one can generate a class activation map (CAM) \cite{zhou2016learning}, with which we can localize the crucial regions of a 2D matrix for a CNN to succeed in classification tasks. Based on the CAM method, R.R.Selvaraju et al. proposed a new CNN interpretation method called Gradient-weighted Class Activation Mapping (Grad-CAM) \cite{selvaraju2017grad}. Compared with the former one, Grad-CAM can be applicable to more types of CNNs and can give more information about what the neuron network learns.

In the CAM method, the class activation map for class $c$ (in our case, $c$ is trivial and can be supressed since we treat a regression instead of a classification problem) is defined as ${M_{c}}$:
\begin{equation}
   M_{c}(x, y) = \sum\limits_k \omega_k^c f_k(x,y).
\end{equation}
Here, $f_k(x,y)$ represents the activation of unit $k$ in the last convolutional layer at spatial location $(x,y)$ and $\omega_k^c$ is the weight measuring the importance of unit $k$ for class $c$. In Grad-CAM, the $\omega_k^c$ is defined as the result of performing global average pooling to the gradient of the score for class $c$ with respect to the activations $f_k$:
\begin{equation}
   \omega_k^c = \frac{1}{Z} \sum\limits_x \sum\limits_y \frac{\partial y^c}{\partial f_k(x,y)},
\end{equation}
where $(1/Z)\sum\limits_x \sum\limits_y$ means the operation of global average pooling. Then the gradient-weighted class activation map is given as:
\begin{equation}
   L^c = ReLU \left(\sum_k \omega_k^c f^k \right).
\end{equation}

CAM and Grad-CAM have succeeded in classification problems. However, we need some adjustments in the definitions of the maps in order to apply them to our CNN for regression. When Grad-CAM meets classification tasks, only the regions correlated positively with the class of interest should be preserved. Thus, a ReLU operation is performed on the linear combination of maps. But for regression problems, both the positively-correlated and the negatively-correlated features ought to be considered. Consequently, we redefine an activation map as the absolute value of the linear combination of maps:
\begin{equation}
L^c = Abs \left(\sum_k \omega_k^c f^k \right).
\end{equation}

\begin{figure}[h]
	\centering
	\includegraphics[width=0.5\textwidth]{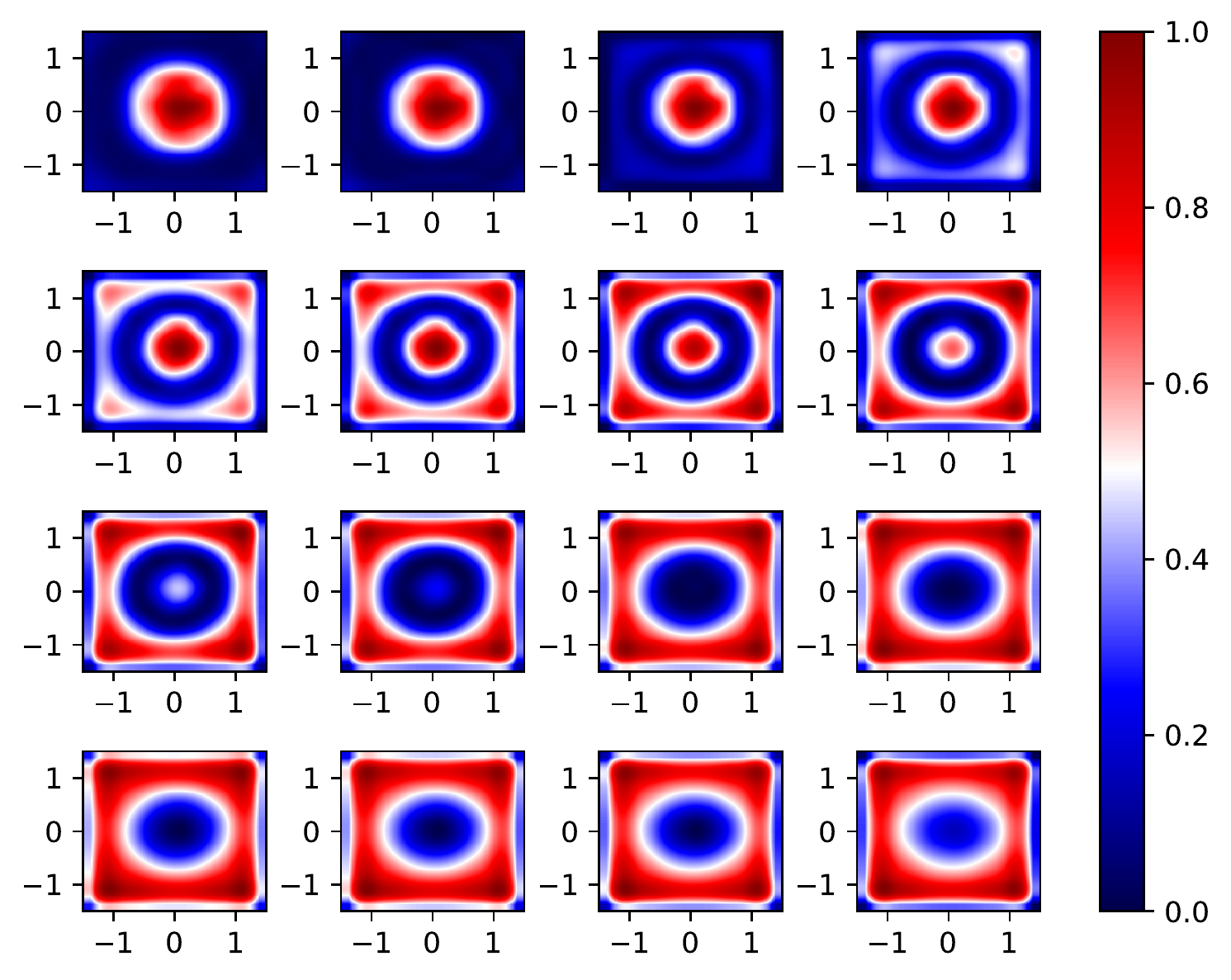}
	\caption{The average `attention' maps for the cases of $b$ = 1.65, 3.54, 4.69, 5.07, 5.44, 5.79, 6.09, 6.41, 6.77, 7.19, 7.64, 8.47, 8.89, 9.30, 9.80, 10.35, 11.1 fm.}
	\label{p11}
\end{figure}

We obtain average `attention' maps for impact parameters on the interval $[2, 11]$ fm where our CNN behaves well in prediction (see Fig.\ref{p11}). It turns out that compared with the cases of central collision, the CNN turns its `attention' to the regions of larger transverse momentum for peripheral collisions. The CNN gives high marks to charged particles with small transverse momentum when it tries to `recognize' a central event's impact parameter. With the increase of $b$, the peripheral area of the energy spectrum begins to attract the CNN's attention and becomes more and more important for distinguishing peripheral events from central ones, which means, the CNN focuses on the particles with larger transverse momentum for peripheral events. 

\section{Summary}\label{sum}
In this paper, we investigate the feasibility of applying the method of deep learning to determining a single event's impact parameter in relativistic heavy-ion collisions. By constructing proper DNN model and CNN model, we establish a map between the energy spectrum of final-state charged particles and the impact parameter for a single collision event.

Simulated by the AMPT model, 180,000 events of Au + Au collision at $\sqrt{s_{NN}} = 200\ GeV$ are generated for training the DL models. In addition, 72,000 events compose the validation dataset aiming at measuring the performance of the DL models. After selecting the best one from the trained models, 20,000 events are used to test its prediction accuracy. The DNN model and CNN model all show high accuracy in the cases of semi-central and semi-peripheral collisions, i.e. with impact parameter $2-10$ fm. The mean prediction error of the CNN model for events with $2 \leq b \leq 11$ fm ranges from -0.06 fm to 0.05 fm and that of the DNN model ranges from -0.08 fm to 0.08 fm. The mean absolute prediction error on the $b$ interval $[2, 12.5]$ fm is 0.40 $fm$ for the CNN model. But these two DL models work worse for central and peripheral collisions.

To investigate the influence of beam energy in this task, we apply these two DL models with the same architectures to the cases of $\sqrt{s_{NN}} = 7.7,\ 39.0,\ 62.4,\ 130.0$ as well. The DL models turn out to be effective for both low energy and high energy collisions. By extending the pseudorapidity cut, the performance of the CNN is improved. And now it reconstructs the impact parameter satisfying $2 \leq b \leq 12.5$ fm with the mean absolute error 0.35 fm. Then, we propose a scheme to compare the uncertainty of $b$ determination by multiplicity method with the CNN method. The charged-particle energy spectrum is proved to be a better observable than the multiplicity to realize this kind of impact parameter determination.

By modifying the Grad-CAM algorithm for classification tasks to the one available for regression tasks, we obtain the `attention' maps for the CNN model. When meeting central collisions, the CNN has a tendency to focus on the particles with small transverse momentum and the situation is opposite for peripheral events.

\section*{Acknowledgement}
We acknowledge useful discussions with W. B. He, L. G. Pang, X. N. Wang, and K. Zhou. The work is supported by NSFC through Grant No.~12075061 and Shanghai NSF through Grant No.~20ZR1404100.


\newpage

\bibliography{refs}

\end{document}